\documentclass{article}
\usepackage{natbib} 					
\usepackage{graphicx}
\usepackage[scriptsize]{caption}
\usepackage{amsfonts}
\usepackage{amsmath}
\usepackage{lscape}
\usepackage{booktabs}

\usepackage[colorlinks]{hyperref} 	
\hypersetup{  
citecolor=black, 						
linkcolor= black} 						

\begin{document}
\title{On Hurst exponent estimation under heavy-tailed distributions}
\date{ }
 \author{
 Jozef Barunik \thanks{%
 Institute of Information Theory and Automation, Academy of Sciences of the Czech Republic, Institute of Economic Studies, Charles University, Prague, Email: barunik@utia.cas.cz (\textit{Corresponding Author})}\hspace{10 mm}  
 Ladislav Kristoufek \thanks{%
 Institute of Information Theory and Automation, Academy of Sciences of the Czech Republic, Institute of Economic Studies, Charles University, Prague}
 }
 
\maketitle

\begin{abstract}
In this paper, we show how the sampling properties of Hurst exponent methods of estimation change with the presence of heavy tails. We run extensive Monte Carlo simulations to find out how rescaled range analysis ($R/S$), multifractal detrended fluctuation analysis ($MF-DFA$), detrending moving average ($DMA$) and generalized Hurst exponent approach ($GHE$) estimate Hurst exponent on independent series with different heavy tails.  For this purpose, we generate independent random series from stable distribution with stability exponent $\alpha$ changing from 1.1 (heaviest tails) to 2 (Gaussian normal distribution) and we estimate Hurst exponent using the different methods. $R/S$ and $GHE$ prove to be robust to heavy tails in the underlying process. $GHE$ provides the lowest variance and bias in comparison to the other methods regardless the presence of heavy tails in data and sample size. Utilizing this result, we apply a novel approach of intraday time-dependent Hurst exponent and we estimate Hurst exponent on high frequency data for each trading day separately. We obtain Hurst exponents for S\&P500 index for the period beginning with year 1983 and ending by November 2009 and we discuss surprising result which uncovers how the market's behavior changed over this long period. \\

\end{abstract}

\newpage
\section{Introduction}
The presence of long-range dependence and the estimation of its characteristic Hurst exponent in financial time series have been widely discussed in several recent research papers, e.g. \citep{alvaetal2008,carboneetal2004,czarnecki2008,dimatteo2007,dimatteoetal2005,grech2004,matos2008,peters1994,Vandewalle1997}. Even though there are several studies of the finite sample properties of different methods \citep{couillarddavidson2005,ellis2007,grech2005,weron2002}, most authors still interpret the results only on the basis of a simple comparison of the estimated Hurst exponent $H$ with the theoretical value for an independent process of 0.5. In more detail, Hurst exponent of 0.5 indicates two possible processes -- either an independent one \citep{beran} or a short-range dependent one \citep{lillofarmer2004}. If $H > 0.5$, the process has significantly positive and non-summable auto-covariance coefficients at all lags and it is called persistent \citep{mandelbrot1968}. When $H < 0.5$, the process has significantly negative summable auto-covariances at all lags and the process is said to be anti-persistent \citep{barkoulasetal2000}. For the persistent process, an increment (decrement) is statistically more likely to be followed by another increment (decrement). Reversely, the anti-persistent process reverts more frequently than the independent process. However, the estimates for a random process can strongly deviate from the limit of $0.5$ \citep{couillarddavidson2005,weron2002}. 

Most authors estimate expected values and confidence intervals based on simulations of a standard normal process $N(0,1)$. The returns of the financial markets, however, are not normally distributed \citep{cont2001} and are  heavy-tailed. The validity of such approach can thus be questioned. This paper tries to fill the gap with original results for independent processes drawn from the family of stable distributions. Stable distributions are peaked and fat-tailed for different parameters, which fits the financial market data more closely. Thus, we want to find out how the sampling properties of the Hurst exponent estimates change with fat tails. For this purpose, expected values and confidence intervals are estimated for rescaled range analysis ($R/S$), multifractal detrended fluctuation analysis ($MF-DFA$), detrending moving average ($DMA$) and generalized Hurst exponent approach ($GHE$) for time series of lengths ranging from 512 to 65,536 observations.

The paper is organized as follows. In Section 2, rescaled range analysis, detrended fluctuation analysis, detrending moving average and generalized Hurst exponent are described in detail. Section 3 focuses on the finite sample properties of the different Hurst exponent estimators. Simulation methodology, stable distributions and research design are presented as well as the results of the Monte Carlo study. After examining the finite sample properties of the different Hurst exponent estimation methods, we use these results and we estimate daily and monthly Hurst exponents on high frequency data.

\section{Hurst exponent estimation}
Many methods of the Hurst exponent estimation have been proposed and used in literature and each of these methods has specific advantages and disadvantages. In our research, we focus on classical methods of rescaled range analysis \citep{hurst1951} and detrended fluctuation analysis \citep{pengetal1994} as well as quite novel methods of multifractal detrended fluctuation analysis \citep{Kantelhardt2002}, detrending moving average \citep{Alessio2002} and generalized Hurst exponent approach \citep{dimatteo2003}. In this section, we shortly describe each method and we follow with a brief review of papers dealing with the finite sample properties of the estimators.

\subsection{Rescaled Range Analysis}
Rescaled range analysis $(R/S)$ was developed by Harold E. Hurst while working as a water engineer in Egypt \citep{hurst1951} and further augmented by \citep{mandelbrotwallis1968}. In the procedure, one takes the returns of the time series of length $T$ and divides them into $N$ adjacent sub-periods of length $\upsilon$, where $N\upsilon=T$. For each sub-period, one calculates the average value and constructs new series of accumulated deviations from the arithmetic mean values (the profile). Then the range, which is defined as the difference between the maximum and the minimum value of the profile, and the standard deviation of the original time series for each sub-period, is calculated. Each range is standardized by the corresponding standard deviation and forms a rescaled range so that the average rescaled range for a given sub-period of length $(R/S)_{\upsilon}$ is calculated. The rescaled range scales as $(R/S)_{\upsilon}\approx c\upsilon^{H}$, where $c$ is a finite constant independent of $\upsilon$ \citep{dimatteo2007,taqquetal1995}. To uncover the scaling law, simple ordinary least squares regression on logarithms of each side is used so that $\log{(R/S)}_{\upsilon}\approx\log{c}+H\log{\upsilon}$, where $H$ is the Hurst exponent.

$R/S$ was shown to be biased for small $\upsilon$ (e.g. \citep{couillarddavidson2005}). The expected values of rescaled ranges for finite samples are given in \cite{Anis1976}. For a discussion of $R/S$ variance, see a recent paper of \cite{Mielniczuk2007}. In our research, we use the original version described in this section.
 
 \subsection{Multifractal Detrended Fluctuation Analysis}
\label{sec2.2}
Multifractal detrended fluctuation analysis ($MF-DFA$), proposed by \citep{Kantelhardt2002}, is a generalization of detrended fluctuation analysis ($DFA$), proposed by \citep{pengetal1994}. The main advantage of this technique is that it can be used for non-stationary multifractal data. Basic notion is the examination of deviations from polynomial fit of different moments $q$. For $q=2$, $MF-DFA$ turns into standard $DFA$.

In the procedure, the time series is divided into sub-periods and the profile is constructed in the same way as for rescaled range analysis. The polynomial fit of order $l$, labeled as $X_{\upsilon,l}$, of the profile is estimated for each sub-period. In our analysis, we set $l=1$ and thus we use the linear detrending. Therefore, we omit the $l$ labeling in the rest of the text. A detrended signal $Y(t,i)$, for sub-period $i=1,\ldots,N$ is then constructed as $Y(t,i)=X(t,i)-X_{\upsilon}(t,i)$. Fluctuation $F^2_{DFA,q}(\upsilon,i)$, which is defined for each sub-period of length $\upsilon$ as $F^2_{DFA,q}(i)=(1/\upsilon\sum_{i=1}^{\upsilon}{Y(t,i)^{2})^{q/2}}$, is then averaged over $N$ sub-periods of length $\upsilon$ for different $q$ and forms $F_{DFA,q}(\upsilon)=(1/N\sum_{i=1}^{N}{F^2_{DFA,q}(i)})^{1/q}$ which then scales as $F_{DFA,q}(\upsilon)\approx c\upsilon^{H(q)}$, where again $c$ is a constant independent of $\upsilon$ and $H(q)$ is the generalized Hurst exponent. The Hurst exponent is again estimated through ordinary least squares regression on logarithms of both sides so that $\log{F_{DFA,q}(\upsilon)}\approx\log{c}+H(q)\log{\upsilon}$ \citep{Kantelhardt2002}.
 
Moreover, $MF-DFA$ has some characteristics which are important for the examination of processes with heavy tails. \citep{Kantelhardt2002} showed that $H(q) \approx 1/q$ for $q>\alpha$ and $H(q) \approx 1/\alpha$ for $q\leq\alpha$, where $\alpha$ is a parameter of stable distributions defined later in the text.
 
\subsection{Detrending Moving Average}

Detrending moving average ($DMA$) is quite a new technique proposed by \citep{Alessio2002} and further discussed in \citep{carboneetal2004}. In the method, one does not need to divide the time series into sub-periods. The method is based on deviations from the moving average of the whole series. Fluctuation $F^2_{DMA, \lambda}$ is defined as $F^2_{DMA, \lambda}=\sum_{t=\lambda}^{T}{(X(t)-\bar{X}_{\lambda}(t))^2}$ where $\bar{X}_{\lambda}(t)=\sum_{k=0}^{\lambda-1}{X(t-\lambda)}$ is the moving average with a time window $\lambda$. The fluctuation then scales as $F^2_{DMA, \lambda}\approx c\lambda^{2H}$.

Moving average $\bar{X}_{\lambda}(t)$ can take various forms in centering (e.g. backward, forward, centered) and weighting (e.g. weighted, not weighted, exponential). For our purposes, we use simple backward moving average. For discussion, see \citep{Alessio2002}.

\subsection{Generalized Hurst Exponent Approach}

Last method used in this study, generalized Hurst exponent approach ($GHE$), is another method suitable for detection of multifractality. The method, recently re-explored for the financial time series by \citep{dimatteo2003}, is based on scaling of $q$-th order moments of the increments of the process $X(t)$. The scaling is characterized on the basis of the statistic $K_q(\tau)$, which is defined as $K_q(\tau)=\sum_{t=0}^{T-\tau}|X(t+\tau)-X(t)|^q/(T-\tau+1)$ for time series of length $T$. The statistic scales as $K_q(\tau)\approx c\tau^{qH(q)}$.

For the purposes of long-range dependence detection, the case of $q=2$ is important as $K_2(\tau)$ is connected to the scaling of the autocorrelation function of the increments. Therefore, we can estimate the Hurst exponent $H(2)$ from relationship $K_q(\tau)\approx c\tau^{2H(2)}$, which is comparable with estimates of $H$ of $R/S$ and $DMA$ and $H(2)$ of $MF-DFA$. For $q=1$, $H(1)$ characterizes the scaling of the absolute deviations of the process \citep{dimatteo2007}.

\subsection{Literature review}

Number of applied papers strongly outnumbers the studies dealing with finite sample properties of the Hurst exponent estimators. However, majority of these studies show that the estimators can differ significantly from the expected Hurst exponent even for independent process. Confidence intervals for testing of the null hypothesis of no long-range dependence are then quite wide. Short chronological review of the studies follows.

\citep{taqquetal1995} presented results of quite wide range of the Hurst exponent estimators (9 in total) on the simulated series of 10,000 observations and 50 realizations. We mention only the results for $R/S$ and $DFA$ concerned by our study. \citep{taqquetal1995} finds that $R/S$ overestimates the true Hurst exponent while $DFA$ underestimates it. Nevertheless, the study is quite limited as it provides small number of simulations for only one set of observations, which is considered to be high for financial applications as 10,000 daily observations equals to approximately 40 years of data.

Extending \citep{taqquetal1995} findings, \citep{Taqqu1998} tested 8 different estimators under different heavy-tailed distributions as well as FARIMA models with heavy-tailed innovations. The authors came to the conclusion that $R/S$ and $DFA$ are quite robust to different distributions whereas both are sensitive to the presence of short-range dependence in the process. However, there are several questions about the methods used and most importantly, the results for $DFA$ need more focus. Firstly, the method is called "variance of residuals method" as is the case for \citep{taqquetal1995}. Secondly, the method is defined in a different way than in \citep{pengetal1994} and the authors claim that for an independent process with stable innovations, the expected value of Hurst exponent is not equal to 0.5 but to $1/\alpha$. However, \citep{Kantelhardt2002} showed that this is the case for $MF-DFA$ with $q=1$ for $1\le\alpha\le2$. \citep{taqquetal1995} is the case as well. Finally, the results are again based on 50 simulations of the length of 10,000.

\citep{weron2002} compares the efficiency of $R/S$ and $DFA$ for an independent underlying process. The author finds that $DFA$ outperforms $R/S$ and also discusses the influence of minimal scales $\upsilon$. The simulations are based on the standard normal distribution. However, the author, similarly to the previous two references, defines the algorithm of $DFA$ as $MF-DFA$ with $q=1$ hence the results must be taken with caution again.

As a generalization of $DFA$, \citep{Kantelhardt2002} proposed its multifractal version ($MF-DFA$) and showed some finite samples results. Most importantly, the authors showed the theoretical relationship between $q$, $\alpha$ and $H$, which is described in Section \ref{sec2.2}. Moreover, the deviations from the true values of generalized Hurst exponent $H(q)$ are shown for both unifractal ($H(q)=H$) and multifractal time series ($H(q)$ varies with $q$).

\citep{couillarddavidson2005} focus on $R/S$ solely and show that the estimates based on \citep{Anis1976} are better than the ones of \citep{peters1994}. The authors show the results on both the rescaled ranges and the Hurst exponent estimates based on the standard normal distribution. Moreover, the authors show that the confidence intervals should not be based on the Central Limit Theorem as the standard deviations of the estimates do not obey the square root law. Such result was already discussed in \citep{weron2002} where the confidence intervals are based on percentiles rather than standard deviations.

\citep{grech2005} compare the finite sample properties of $DFA$ and $DMA$ for the standard normal distribution process. The authors show that $DFA$ outperforms $DMA$ in both bias and variance for the time series of up to 30,000 observations. However, the results are quite questionable due to several issues. Firstly, the estimates are taken into consideration only in the case of $R^2>0.98$ for the Hurst exponent estimation regression without any discussion of a ratio of omitted estimates. Secondly, the choice of lags $\lambda$ is not discussed at all for $DMA$.

The other aspect of $R/S$ is discussed in \citep{ellis2007} who examines the efficiency of $R/S$ method with the use of either contiguous or overlapping sub-series and concludes that the methods do not differ significantly while the confidence intervals show the similar deviation from the Central Limit Theorem as shown in \citep{weron2002}.

Finally, a further research on $R/S$ versus $DFA$ are discussed in \citep{Mielniczuk2007}. The authors show that if the bias of $R/S$ is taken into consideration, the root mean square errors are smaller for $R/S$ than for $DFA$ for all time series lengths from $2^9$ to $2^{15}$ as well as for Hurst exponents from 0.5 to 0.9.

In the following section, we challenge these studies and run the extensive Monte Carlo study to find out how rescaled range analysis ($R/S$), multi-fractal detrended fluctuation analysis ($MF-DFA$), detrending moving average ($DMA$) and generalized Hurst exponent approach ($GHE$) behave when estimating Hurst exponent on independent series with different heavy tails and different time series lengths ranging from 512 to 65,536 observations.

\section{Finite sample properties of different Hurst exponent estimators}
The primary objective of this paper is to examine the robustness of the various Hurst exponent estimation methods to heavy-tailed data and contribute to the previous discussions. 

The majority of the research papers dealing with the finite sample properties of Hurst exponent estimation methods base their results on simulations of the standardized normal distribution $N(0,1)$. However, a wide spectrum of real-world data is non-normal so that we cannot really use these results for statistical inference when estimating Hurst exponent on these data. To examine the behavior of the estimators of Hurst exponent for an $i.i.d.$ non-normal process with heavy tails, we take advantage of changing the parameter $\alpha$ in the stable distributions. Before we continue in describing the research design, we briefly introduce the stable distributions.

\subsection{Stable distributions}
Stable distributions form a class of probability laws with appealing theoretical properties which fit  well-known stylized facts about financial markets such as skewness and excess kurtosis. Such distributions are described by four parameters -- $\alpha,\beta,\gamma,\delta$. $\alpha$ is the characteristic exponent and $0<\alpha \leq2\ $. $\beta$ is the skewness parameter and $-1 \leq \beta \leq 1$. For $\beta = 0$, the distribution is symmetric; for $\beta > 0$, it is skewed to the right; and for $\beta < 0$, it is skewed to the left. While the parameters $\alpha$ and $\beta$ determine the shape of the distribution, $\gamma$ and $\delta$ are scale and location parameters, respectively.

Due to a lack of closed-form formulas for the probability density functions (except for three stable distributions -- Gaussian, Cauchy, Levy), the $\alpha$-stable distribution can be described by a characteristic function which is the inverse Fourier transform of the probability density function, i.e., $\phi(u)=E\exp(iuX)$. In our paper, we use the following characteristic function:
\begin{equation}
 \footnotesize
\label{eq2}
\phi(u)=
\left\{
\begin{array}{lr}
\exp(-\gamma^{\alpha}|u|^{\alpha}[1+i\beta(\tan\frac{\pi\alpha}{2})($sign$ u)(|\gamma u|^{1-\alpha}-1)]+i\delta u) & \alpha \ne 1 \\
\exp(-\gamma |u| [1+i\beta\frac{2}{\pi}($sign$ u) \ln (\gamma |u|)]+i\delta u) & \alpha = 1
\end{array}
\right.\end{equation}
In order to simulate random stable variables, we use a method of \cite{Chambers1976} which is more efficient than inverse Fourier Transforms method \citep{Menn2006} and was proved to be accurate by \cite{Weron1996}.  

For all values of the parameters $\alpha<2$ and $-1<\beta<1$, the stable distributions have two tails that are asymptotically power laws. If the data come from stable distribution, the empirical distribution function should be approximately a straight line with slope $-\alpha$ in a log-log plot.

For the financial market analysis, the characteristic exponent $\alpha$ gives additional important information. When $\alpha < 2$, extreme events are more probable than for the Gaussian distribution ($\alpha=2$). From an economic point of view, $0 < \alpha < 1$ implies infinite expected returns. To avoid this, we examine only the cases when $1 < \alpha \le 2$.

For more details about stable distributions, we refer readers to \citep{Zoltarev, SamorodintskyTaquu1994,Borak2005}.

\subsection{Monte Carlo design}
The simulations are constructed in order to find out how $R/S$, $DFA$, $MF-DFA$, $DMA$ and $GHE$  methods react to non-normal data with heavy tails. We use $i.i.d.$ $\alpha$-stable distributed random variables from $S(\alpha,0,\sqrt{2}/2,0)$, where $1.1 \le \alpha \le 2$ with a step of 0.1. For each parameter $\alpha$, 1,000 time series with lengths from $2^{9}$ to $2^{16}$ are simulated and Hurst exponents for $R/S$, $DFA$ (eqivalent to $MF-DFA$ with $q=2$), $MF-DFA$ with $q=1$, $DMA$, $GHE$ with $q=1$ and $GHE$ with $q=2$ are estimated.

In other words, we simulate the grid of independent identically distributed stable increments with the different $\alpha$ for various series lengths. For each position in the grid, we estimate Hurst exponents with the use of the different methods. This allows us to see how the finite sample properties of the estimation methods change with the varying coefficient $\alpha$. The first set, with $\alpha=1.1$, has the heaviest tails, and with increasing $\alpha$, the tails approach the Gaussian normal distribution with $\alpha=2$. Moreover, each set shows how the estimates behave with changing time series lengths.

\begin{figure}
   \centering
   \includegraphics[width=5in]{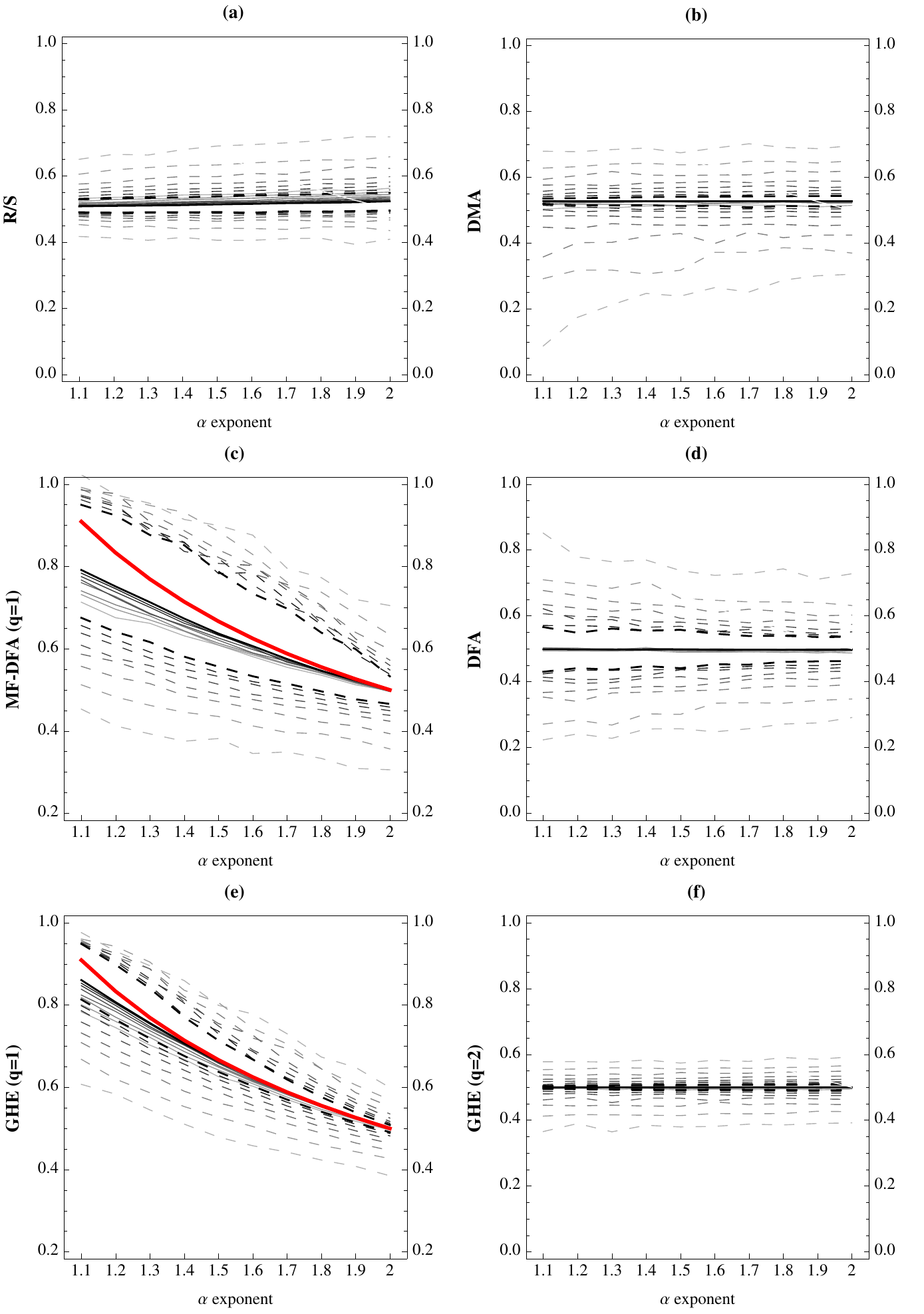} 
  \caption{Expected values of $H$ with 2.5\% and 97.5\% quantiles for the different $\alpha$s and methods of estimation. Figures show the results for (a) $R/S$, (b) $DMA$, (c) $MF-DFA$ with $q=1$, (d) $DFA$,(e) $GHE$ for $q=1$ and (f) $GHE$ for $q=2$. The horizontal axis goes from simulations with the heaviest tails ($\alpha=1.1$) to the simulations with the normal distribution ($\alpha=2$). Figures also show results for different simulated time series lengths from $2^{9}$ to $2^{16}$ with the longest time series of $2^{16}$ plotted in the boldest black. The lower the time series length, the greyer the plots. The mean values are plotted as full lines while the quantiles are dashed lines. Figures (c) and (e) also contain the theoretical expected values of $H(q) \approx 1/\alpha$ for $q\leq\alpha$ in bold red. (The theoretical expected value of the other methods is 0.5.)}
   \label{fig:ex2}
\end{figure}

Table \ref{tab1} summarizes the expected values of the estimated Hurst exponents for the grid. Table \ref{tab2} and Table \ref{tab3} show the 2.5\% and the 97.5\% quantiles of the Hurst exponent estimates. Figure \ref{fig:ex2} illustrates all the results and serves best for the quick orientation. It shows the estimated expected values with the 2.5\% and the 97.5\% quantiles for all time series lengths and for all methods considered by our Monte Carlo simulation. We discuss the results for each method separately. 

\subsection{Results for $R/S$}
We can see that the expected value of the $R/S$ method\footnote{For the estimations of $R/S$ and $MF-DFA$, we use a sub-length $\upsilon$ equal to the power of a set integer value \citep{weron2002}. Thus, we set a basis $b$, a minimum power $pmin$, and a maximum power $pmax$ so that we get $\upsilon=b^{pmin},\ldots,b^{pmax}$, where $b^{pmin}\ge b$ is the minimum scale and $b^{pmax}\le T$ is the maximum scale. The minimum scale is set to 16 observations and the maximum one is set as a quarter of the time series length according to \citep{alvarezetal2005,einstein2001,grech2004,matos2008,peters1994}. Such choice avoids biases of low scales, where standard deviations and trends can be estimated inefficiently, and high scales, where the extreme values need not be averaged out. Moreover, the used intervals are contingent and non-overlapping -- see \citep{ellis2007} for discussion.} is converging to 0.5 with heavier tails. For length $2^{9}$, the expected value of $R/S$ (with quantiles in brackets) for the simulations with heaviest tails ($\alpha=1.1$) is 0.5318 (0.4172, 0.6502) while for length $2^{16}$, the expected value (quantiles) of $R/S$ with the same $\alpha=1.1$ is 0.5082 (0.4904, 0.5291). With growing $\alpha$, the expected value of $R/S$ also grows slowly. For the time series of length $2^{9}$, the expected value of $R/S$ for the simulations with the Gaussian normal distribution ($\alpha=2$) is 0.5627 (0.4088, 0.7179) while for length $2^{16}$, the mean of $R/S$ with the same $\alpha=2$ is 0.5235 (0.4956, 0.5489). Tables \ref{tab1}, \ref{tab2}, and \ref{tab3} together with Figure \ref{fig:ex2} (a) give the full picture. We can see that the length of the simulated time series improves the estimates through all $\alpha$s. Moreover, the expected value of the estimated $H$ is slowly converging to 0.5 with $\alpha$ changing from 2 to 1.1.

$R/S$ method is thus quite robust to the presence of heavy tails in the underlying data. It is interesting that the expected value of $H$ is converging to 0.5 with decreasing value of $\alpha$. In other words, the heavier the tails in the data are, the closer the expected estimate of $H$ to the value of 0.5 is. Moreover, the expected value of $H$ is slowly converging to 0.5 with increasing length of the sample.

\subsection{Results for $DMA$}
$DMA$ has very similar finite sample properties to $R/S$. The difference is in wider confidence intervals for small samples for DMA. On the contrary, $DMA$ has narrower confidence intervals for longer time series which makes it more precise. For the series with the heaviest tails ($\alpha=1.1$) and length $2^{9}$, the expected value (quantiles) of $DMA$ is 0.5036 (0.0877, 0.6792 ) while for length $2^{16}$, the mean (quantiles) of $DMA$ with the same $\alpha=1.1$ is 0.5276 (0.5192, 0.5358). With $\alpha$ approaching the value of 2, which represents the Gaussian normal distribution, the expected value of $DMA$ does not significantly change. More precisely, for the time series of length $2^{9}$, the expected value of $DMA$ for the simulations with the lightest tails ($\alpha=2$) is 0.5147 (0.3054, 0.6959) while for length $2^{16}$, the mean of $DMA$ with the same $\alpha=2$ is 0.5272 (0.5092, 0.5436). Again, Tables \ref{tab1}, \ref{tab2}, and \ref{tab3} together with Figure \ref{fig:ex2} (b) give the full picture. We have to make a note that we tried to estimate DMA with different window lengths $\lambda$ and minimums $\lambda_{min}$ as the method is very sensitive to these parameters. Finally, we set $\lambda_{min}=20$ and $\lambda_{max}=40$, which may produce bias when being used on larger datasets. But in order to maintain the comparative power, we choose to set the parameters for all simulations.

\subsection{Results for $DFA$}
Let's move to $DFA$ method, which seems to be even less precise than the previous methods of estimation. For the series with the heaviest tails ($\alpha=1.1$) and length $2^{9}$, the expected value (quantiles) of $DFA$ is 0.5027 (0.2234, 0.8513) while for length $2^{16}$, the mean (quantiles) of $DFA$ with the same $\alpha=1.1$ is 0.4984 (0.4301, 0.5659 ). With $\alpha$ approaching the value of 2, which represents the Gaussian normal distribution, the expected value of $DFA$ is still very close to 0.5 and it has slightly narrower confidence intervals. More precisely, for the time series of length $2^{9}$, the expected value of $DFA$ for the Gaussian normal distribution ($\alpha=2$) is 0.4940 (0.2910, 0.7280) while for length $2^{16}$, the mean of $DFA$ with the same $\alpha=2$ is 0.4981 (0.4630, 0.5361). See Tables \ref{tab1}, \ref{tab2}, and \ref{tab3} together with Figure \ref{fig:ex2} (d) for the full picture. Altogether, $DFA$ yields the closest expected estimate of $H$ for all different $\alpha$ exponents and for all lengths. Unfortunately, much wider confidence intervals make any statistical conclusions, especially for small samples, hard to state.

\subsection{Results for $MF-DFA$}
We simulated more general $MF-DFA$ method with $q=1$ so that the expected value of $H$ is $1/\alpha$ according to \citep{Kantelhardt2002} (In fact, the previously discussed $DFA$ is the special case of $MF-DFA$ with $q=2$). Results show that its expected value is diverging from 0.5 with heavier tails as expected but it underestimates the true value. For the series with the heaviest tails ($\alpha=1.1$) and length $2^{9}$, the expected value (quantiles) of $MF-DFA(q=1)$ is 0.7132 (0.4533, 1.0230 ) while for length $2^{16}$, the mean (quantiles) of $MF-DFA(q=1)$ with the same $\alpha=1.1$ is 0.7915 (0.6752, 0.9499). With $\alpha$ approaching the value of 2, which represents the Gaussian normal distribution, the expected value of $MF-DFA(q=1)$ approaches 0.5. More precisely, for the time series of length $2^{9}$, the expected value of $MF-DFA(q=1)$  for the simulations with the lightest tails ($\alpha=2$) is 0.4973 (0.3059, 0.7280) while for length $2^{16}$, the mean of $MF-DFA(q=1)$  with the same $\alpha=2$ is 0.4996 (0.4659, 0.5330). See Tables \ref{tab1}, \ref{tab2}, and \ref{tab3} together with Figure \ref{fig:ex2} (c) for the full picture. We can see that even  $MF-DFA(q=1)$ is quite inaccurate method for all $\alpha$ exponents and all series lengths (comparable to $MF-DFA(q=2)$ which is equivalent to $DFA$).

\subsection{Results for $GHE$}
Finally, we turn to $GHE$ methods. Expected value of $GHE(q=1)$ is again $1/\alpha$ as in the case of $MF-DFA(q=1)$. For length $2^{9}$, the expected value of $GHE(q=1)$ (with quantiles in brackets) for the simulations with the heaviest tails ($\alpha=1.1$) is 0.7832 (0.6072, 0.9761) while for length $2^{16}$, the mean (quantiles) of $GHE(q=1)$ with the same $\alpha=1.1$ is 0.8607 (0.8134, 0.9488). Thus $GHE(q=1)$ yields considerably better results than $MF-DFA(q=1)$ as it provides much narrower confidence intervals and it does not underestimate the expected value so strongly under the presence of heavier tails. For the time series of length $2^{9}$, the expected value of $GHE(q=1)$ for the simulations with the lightest tails ($\alpha=2$) is 0.4963 (0.3850, 0.5993) while for length $2^{16}$, the mean of $GHE(q=1)$ with the same $\alpha=2$ is 0.5001 (0.4910, 0.5085). Tables \ref{tab1}, \ref{tab2}, and \ref{tab3} together with Figure \ref{fig:ex2} (e) give the full picture. We can see that $GHE(q=1)$ yields much better results that $MF-DFA(q=1)$.

For length $2^{9}$, the expected value of $GHE(q=2)$ (with quantiles in brackets) for the simulations with the heaviest tails ($\alpha=1.1$) is 0.4955 (0.3657, 0.5777) while for length $2^{16}$, the mean (quantiles) of $GHE(q=2)$ with the same $\alpha=1.1$ is 0.4999 (.4963, 0.5036). Thus $GHE(q=2)$ yields considerably better results than other methods as it provides much narrower confidence intervals. For the time series of length $2^{9}$, the expected value of $GHE(q=2)$ for the simulations with the Gaussian normal distribution ($\alpha=2$) is 0.4946 (0.3920, 0.5926) while for length $2^{16}$, the mean of $GHE(q=2)$ with the same $\alpha=2$ is 0.5000 (0.4915, 0.5082). Tables \ref{tab1}, \ref{tab2}, and \ref{tab3} together with Figure \ref{fig:ex2} (f) give the full picture. We can see that $GHE$ with $q=2$ provides the best finite sample behavior among all the methods in the means of the lowest variance and bias.

\subsection{Comments on Results}
\begin{landscape}
\begin{figure}[h]
   \centering
   \includegraphics[width=7in]{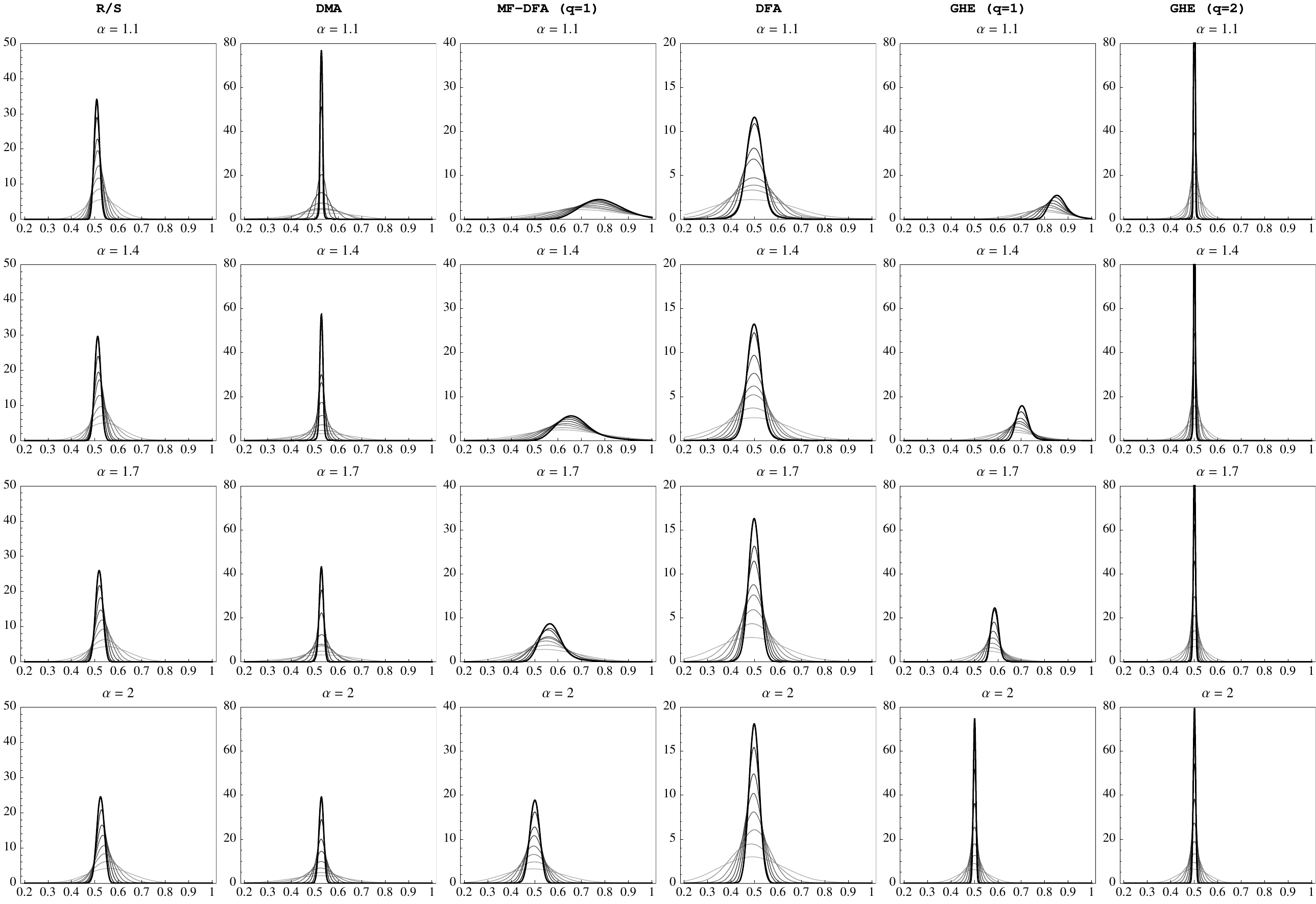} 
  \caption{Probability distribution functions of estimated $H$. The grid shows the results with the different methods of estimation ($R/S$, $DMA$, $MF-DFA$ with $q=1$, $DFA$ and $GHE$ for $q=1$ and $q=2$) in the columns and different $\alpha$s in the rows, with the simulations with the heaviest tails in the first row and the simulations with normal distribution in the last row. Moreover, each part of the grid shows 8 lines representing the PDFs of the estimates of $H$ for different simulated time series lengths from $2^{9}$ to $2^{16}$, with the longest time series of $2^{16}$ plotted in the boldest black. The lower the time series length, the greyer the plots.}
   \label{fig:ex0}
\end{figure}
\end{landscape}

We only discussed the results of our simulations with the limiting values of $\alpha$ equal to  1.1 and 2 representing the simulations of time series with the heaviest tails and time series with the tails of the normal distribution, respectively. Tables \ref{tab1}, \ref{tab2}, and \ref{tab3} as well as Figures \ref{fig:ex2} and \ref{fig:ex0} show the whole grid of the simulated results so that the reader can see how the values change for different tail values and can also obtain precise quantile values. Figure \ref{fig:ex0} summarizes all the simulations as it shows the estimated probability distribution functions of all the estimated Hurst exponents using the different methods of estimation for the different values of $\alpha$. For each part of the grid, Figure \ref{fig:ex0} also shows the probability distribution functions for different lengths of simulated time series from $2^{9}$ up to $2^{16}$.

We can see that $R/S$ and $GHE$ with $q=2$ methods are robust to heavy tails. In fact, it is quite interesting that the confidence intervals of the $R/S$ estimate are narrowing with heavier tails. $GHE$ provides the most narrow confidence intervals. Even if we consider the $GHE$ with $q=1$ which has expected value of $H$ equal $1/\alpha$, it holds very good finite sample properties as well. On the other hand, all the other methods ($DMA$, $DFA$, $MF-DFA(q=1)$) deteriorate with heavier tails in the underlying distributions. While all the methods hold the expected value for all time series lengths and thus seem to be better for $H$ estimation than $R/S$ for the normal data with $\alpha=2$, the situation changes considerably for the non-normal simulations. The heavier the tails of the underlying data are, the wider the confidence intervals of estimate are. $MF-DFA(q=1)$ also tends to underestimate the expected $1/\alpha$ value.

In fact, we can see that $MF-DFA$ methods as well as $DMA$ are not appropriate for data with heavier tails and for small sample size. On the other hand, both $GHE$ tested methods proved to be very useful as they outperform all other methods quite strongly.

Note that our results are quite contrary to the ones of \cite{weron2002} who showed that $DFA$ outperforms $R/S$ in both bias and variance. However, the author used $\upsilon_{max}=T$ whereas we use $\upsilon_{max}=T/4$ for both methods. Hence, the choice of $\upsilon_{min}$ and $\upsilon_{max}$ strongly influences the final results of simulations. However, such choice has not been rigorously discussed in the literature yet and it should be of an interest for future research. 

 \section{Empirical study: Application to HF data}
We showed that generalized Hurst exponent approach ($GHE$) can be used regardless the presence of heavy tails in the data and it has by far the best finite sample properties for short sample data. Thus we utilize this result and apply $GHE$ to estimate the intraday time-dependent Hurst exponent of S\&P 500 index on 1-minute data from the period beginning with 2.1.1983 and ending 30.11.2009. In other words, we estimate 6,764 values of the local Hurst exponents while each value is estimated on the information from one day. Thus the Hurst exponents for different days can be compared statistically as the final time series of intraday time-dependent Hurst exponents is not dependent as in the case of time-dependent Hurst exponent using a moving window \citep{grech2004,czarnecki2008}. Each value of the Hurst exponent is estimated for a single day on the intraday data so that we expect all the values to be close to $0.5$. Generalized Hurst exponent approach with its narrow confidence intervals allows us to detect possible periods with significantly higher or lower persistence.
 
Moreover, we also estimate local Hurst exponents from the same data for each month, providing a different view. The length of the time period allows to study how the persistence changed over  323 months. We have up to 8,000 one minute realizations in each month so that these estimates are very accurate.

\begin{figure}[h]
   \centering
   \includegraphics[width=4.5in]{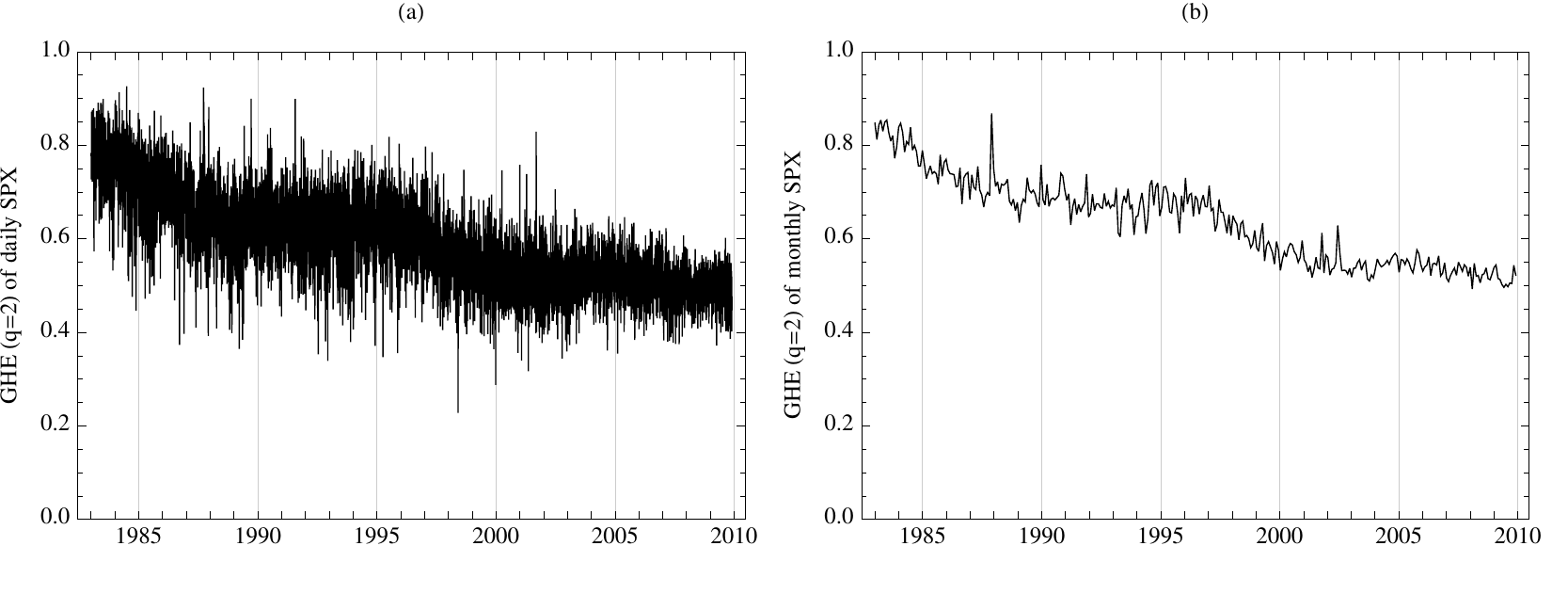} 
  \caption{(a) Hurst exponent estimates using $GHE$ with $q=2$ for each day of 1983 - 2009 period using 1-minute data   (b)  Hurst exponent estimates using $GHE$ with $q=2$ for each month of 1983 - 2009 period using 1-minute data}
   \label{fig:ex1}
\end{figure}

Figure \ref{fig:ex1} shows the results of the estimation for both daily and monthly intraday time-dependent Hurst exponents. It is interesting how Hurst exponent decreases during the period from the level of above $0.8$ to the level of $0.5$. While Hurst exponents on daily data are still noisy, the estimates on monthly series are very strong as confidence intervals of $GHE$ for such long periods of intraday data are very narrow\footnote{For exact confidence intervals, see Tables \ref{tab1}, \ref{tab2}, and \ref{tab3}.}. The Hurst exponents from the last period of 2005-2009 can not be statistically distinguished from 0.5. For the previous periods, H=0.5 can be strongly rejected and we can conclude that the Hurst exponents are statistically different from its theoretical expected value for all the previous periods.

The result indicates how stock market structure changed from 1980s until now. Probably the introduction of algorithmic trading, exponentially rising volume of trades, is the main reason behind this change. Still, the result is quite surprising as we expected the Hurst exponent estimates to show only slight difference from its expected value of 0.5. 

\section{Conclusion}
In this paper, we question the sampling properties of the Hurst exponent estimation methods under heavy-tailed underlying processes. Most existing studies focus on estimating expected values and confidence intervals based on simulations of a standard normal process. However, the returns of the financial markets are not normally distributed, which makes these results impracticable. This paper tries to fill the gap with the original results for independent processes drawn from the family of stable distributions, allowing for heavy tails.

The Monte Carlo study shows that rescaled range analysis ($R/S$) together with generalized Hurst exponent approach ($GHE$) are robust to heavy tails in the underlying process. Detrended moving average ($DMA$) together with detrended fluctuation analysis ($DFA$) and its multifractal generalization ($MF-DFA$) deteriorate with increasing heavy tails in the underlying distributions. While on normal data with $\alpha=2$, all the methods hold the expected value for all time series lengths and thus they seem to be better for $H$ estimation than $R/S$, the situation changes considerably on non-normal simulations. The heavier the tails of the underlying data are, the wider the confidence intervals of the estimates are. $MF-DFA(q=1)$ tends to underestimate the expected $1/\alpha$ value. We can conclude that $MF-DFA$ methods as well as $DMA$ are not appropriate for data with heavier tails and small sample size. On the other hand, both $GHE$ tested methods proved to be very useful as they show the best properties.

Finally, we utilize the results from our simulations on the high frequency data. On the 1-minute S\&P 500 index data from the period of 1983 to 2009, we estimate the generalized Hurst exponent based on $GHE$ for each day as well as for each month. This way, we estimate the Hurst exponents on separate information sets while the finite sample properties of $GHE$ allow us to use it regardless the presence of heavy tails even on smaller samples. The final result is quite surprising as it shows how the U.S. stock market changed from the strongly persistent in 1980s to independent one in the period 2005-2009. 

\section*{Acknowledgements}
The authors would like to thank the anonymous referees whose suggestions helped to improve this paper substantially. The support from the Czech Science Foundation under grant 402/09/0965, support from the Grant Agency of Charles University (GAUK) under projects 118310 and 46108 and Institutional Support from the Department of Education MSMT 0021620841 are gratefully acknowledged.

\bibliography{Physica_A_Hurst}
\bibliographystyle{chicago}

\section*{Figures and Tables}

\begin{table}[c]
\footnotesize
\begin{tabular}{@{ } l@{\hspace{4mm}} l@{\hspace{4mm}}c@{\hspace{2mm}}c@{\hspace{2mm}}c@{\hspace{2mm}}c@{ \hspace{2mm}}c@{\hspace{2mm}}c@{ \hspace{2mm}}c@{\hspace{2mm}}c@{\hspace{2mm}}c@{\hspace{2mm}}c@{\hspace{2mm}}c@{\hspace{2mm}}}
 \toprule \toprule
  & & \text{$\alpha $=1.1} & \text{$\alpha $=1.2} & \text{$\alpha $=1.3} & \text{$\alpha $=1.4} & \text{$\alpha $=1.5} & \text{$\alpha $=1.6} & \text{$\alpha
$=1.7} & \text{$\alpha $=1.8} & \text{$\alpha $=1.9} & \text{$\alpha $=2} \\
 \midrule \midrule
 $2^{9}$ 	& R/S    		 &  0.5318 & 0.5353 & 0.5373 & 0.5425 & 0.5435 & 0.5462 & 0.5519 & 0.5585 & 0.5551 & 0.5627 \\
  		& DMA 		 & 0.5036 & 0.5082 & 0.5070 & 0.5126 & 0.5149 & 0.5148 & 0.5157 & 0.5155 & 0.5101 & 0.5147 \\
  		& MF-DFA		 & 0.7132 & 0.6754 & 0.6619 & 0.6315 & 0.6088 & 0.5808 & 0.5550 & 0.5358 & 0.5124 & 0.4973 \\
		& DFA 		 & 0.5027 & 0.5026 & 0.4955 & 0.5042 & 0.4912 & 0.4897 & 0.4902 & 0.4922 & 0.4867 & 0.4940 \\
		& GHE (q=1) 	 & 0.7832 & 0.7452 & 0.7026 & 0.6683 & 0.6259 & 0.6028 & 0.5705 & 0.5423 & 0.5176 & 0.4963 \\
		& GHE (q=2) 	 & 0.4955 & 0.4985 & 0.4923 & 0.4969 & 0.4916 & 0.4944 & 0.4930 & 0.4952 & 0.4921 & 0.4946 \\	
 \midrule
 $2^{10}$ 	& R/S    		 & 0.5228 & 0.5287 & 0.5308 & 0.5326 & 0.5349 & 0.5398 & 0.5435 & 0.5442 & 0.5475 & 0.5513 \\
  		& DMA 		 & 0.5172 & 0.5177 & 0.5179 & 0.5173 & 0.5146 & 0.5205 & 0.5209 & 0.5249 & 0.5236 & 0.5227 \\
  		& MF-DFA		& 0.7317 & 0.6959 & 0.6700 & 0.6426 & 0.6090 & 0.5854 & 0.5612 & 0.5403 & 0.5161 & 0.4966 \\
		& DFA 		 & 0.4938 & 0.4933 & 0.4927 & 0.4949 & 0.4893 & 0.4897 & 0.4914 & 0.4888 & 0.4904 & 0.4871 \\
		& GHE (q=1) 	 & 0.8008 & 0.7603 & 0.7215 & 0.6768 & 0.6421 & 0.6088 & 0.5763 & 0.5465 & 0.5222 & 0.4975 \\
		& GHE (q=2) 	 &  0.4979 & 0.4967 & 0.4975 & 0.4978 & 0.4958 & 0.4962 & 0.4971 & 0.4969 & 0.4964 & 0.4966 \\	
 \midrule
 $2^{11}$ & R/S    		 &  0.5185 & 0.5207 & 0.5244 & 0.5287 & 0.5300 & 0.5326 & 0.5364 & 0.5391 & 0.5419 & 0.5434 \\
  		& DMA 		 &  0.5185 & 0.5223 & 0.5234 & 0.5253 & 0.5253 & 0.5200 & 0.5255 & 0.5248 & 0.5216 & 0.5220 \\
  		& MF-DFA		 &  0.7408 & 0.7067 & 0.6800 & 0.6436 & 0.6134 & 0.5887 & 0.5629 & 0.5420 & 0.5173 & 0.4960 \\
		& DFA 		 & 0.5032 & 0.5001 & 0.4996 & 0.4971 & 0.4953 & 0.4949 & 0.4950 & 0.4948 & 0.4942 & 0.4978 \\
		& GHE (q=1) 	 & 0.8182 & 0.7705 & 0.7294 & 0.6871 & 0.6492 & 0.6135 & 0.5820 & 0.5520 & 0.5228 & 0.5011 \\
		& GHE (q=2) 	 & 0.4988 & 0.4980 & 0.4990 & 0.4979 & 0.4981 & 0.5005 & 0.4991 & 0.4987 & 0.4980 & 0.5004 \\ 	
 \midrule
 $2^{12}$ 	& R/S    		 & 0.5165 & 0.5184 & 0.5198 & 0.5223 & 0.5244 & 0.5255 & 0.5315 & 0.5328 & 0.5360 & 0.5371 \\
  		& DMA 		 & 0.5240 & 0.5243 & 0.5247 & 0.5260 & 0.5253 & 0.5262 & 0.5248 & 0.5256 & 0.5242 & 0.5248 \\
  		& MF-DFA(1) 	 & 0.7607 & 0.7267 & 0.6861 & 0.6535 & 0.6209 & 0.5900 & 0.5695 & 0.5445 & 0.5200 & 0.4960 \\
		& DFA 		 &  0.5000 & 0.4980 & 0.4966 & 0.4982 & 0.4945 & 0.4926 & 0.4951 & 0.4950 & 0.4943 & 0.4949 \\
		& GHE (q=1) 	 & 0.8242 & 0.7844 & 0.7376 & 0.6954 & 0.6523 & 0.6160 & 0.5841 & 0.5533 & 0.5244 & 0.4994 \\
		& GHE (q=2) 	 & 0.4982 & 0.4993 & 0.4985 & 0.4994 & 0.4982 & 0.4983 & 0.4998 & 0.4996 & 0.4995 & 0.4991 \\
 \midrule
 $2^{13}$ 	& R/S    		 &  0.5127 & 0.5144 & 0.5168 & 0.5195 & 0.5217 & 0.5228 & 0.5256 & 0.5290 & 0.5307 & 0.5337 \\
  		& DMA 		 & 0.5268 & 0.5270 & 0.5281 & 0.5274 & 0.5277 & 0.5252 & 0.5266 & 0.5287 & 0.5266 & 0.5275 \\
  		& MF-DFA		 &  0.7671 & 0.7241 & 0.6844 & 0.6564 & 0.6261 & 0.5982 & 0.5721 & 0.5453 & 0.5190 & 0.4984 \\
		& DFA 		 & 0.4989 & 0.4978 & 0.4970 & 0.4980 & 0.4992 & 0.4950 & 0.4954 & 0.4960 & 0.4941 & 0.4951 \\
		& GHE (q=1) 	 & 0.8384 & 0.7909 & 0.7437 & 0.7025 & 0.6583 & 0.6186 & 0.5850 & 0.5548 & 0.5258 & 0.4997 \\
		& GHE (q=2) 	 &  0.5000 & 0.4999 & 0.4996 & 0.4996 & 0.5000 & 0.4998 & 0.4995 & 0.4997 & 0.4994 & 0.4996 \\
 \midrule
 $2^{14}$ 	& R/S    		 & 0.5109 & 0.5124 & 0.5149 & 0.5167 & 0.5191 & 0.5201 & 0.5224 & 0.5250 & 0.5275 & 0.5299 \\
  		& DMA 		 & 0.5260 & 0.5270 & 0.5274 & 0.5280 & 0.5276 & 0.5268 & 0.5268 & 0.5271 & 0.5270 & 0.5269 \\
  		& MF-DFA		& 0.7754 & 0.7379 & 0.7008 & 0.6620 & 0.6346 & 0.6023 & 0.5674 & 0.5465 & 0.5215 & 0.4988 \\
		& DFA 		 & 0.4992 & 0.4988 & 0.4963 & 0.4958 & 0.4981 & 0.4969 & 0.4981 & 0.4977 & 0.4959 & 0.4959 \\ 
		& GHE (q=1) 	 & 0.8470 & 0.7990 & 0.7476 & 0.7038 & 0.6604 & 0.6226 & 0.5866 & 0.5549 & 0.5252 & 0.5002 \\
		& GHE (q=2) 	 & 0.4997 & 0.4996 & 0.4996 & 0.4997 & 0.4990 & 0.5000 & 0.5003 & 0.4998 & 0.4994 & 0.5003 \\
 \midrule
 $2^{15}$ 	& R/S    		 & 0.5087 & 0.5104 & 0.5122 & 0.5139 & 0.5159 & 0.5169 & 0.5193 & 0.5226 & 0.5243 & 0.5266 \\
  		& DMA 		 & 0.5278 & 0.5278 & 0.5269 & 0.5263 & 0.5263 & 0.5265 & 0.5273 & 0.5272 & 0.5279 & 0.5277 \\
  		& MF-DFA		& 0.7837 & 0.7459 & 0.7060 & 0.6674 & 0.6329 & 0.6063 & 0.5756 & 0.5478 & 0.5221 & 0.4996 \\
		& DFA 		 &  0.4988 & 0.4989 & 0.4990 & 0.4964 & 0.4967 & 0.4992 & 0.4986 & 0.4980 & 0.4966 & 0.4961 \\
		& GHE (q=1) 	 & 0.8531 & 0.8036 & 0.7536 & 0.7053 & 0.6632 & 0.6229 & 0.5872 & 0.5549 & 0.5265 & 0.5001 \\
		& GHE (q=2) 	 & 0.4999 & 0.5003 & 0.5001 & 0.5001 & 0.4998 & 0.5000 & 0.5002 & 0.4999 & 0.5001 & 0.5000 \\
 \midrule
 $2^{16}$ 	& R/S    		 &  0.5082 & 0.5097 & 0.5106 & 0.5119 & 0.5141 & 0.5157 & 0.5179 & 0.5188 & 0.5205 & 0.5235 \\
  		& DMA 		 &  0.5276 & 0.5271 & 0.5272 & 0.5275 & 0.5271 & 0.5272 & 0.5272 & 0.5275 & 0.5272 & 0.5272 \\
  		& MF-DFA 	&  0.7915 & 0.7514 & 0.7132 & 0.6734 & 0.6369 & 0.6065 & 0.5751 & 0.5464 & 0.5236 & 0.4996 \\
		& DFA 		 & 0.4984 & 0.4979 & 0.4979 & 0.4986 & 0.4985 & 0.4976 & 0.4972 & 0.4969 & 0.4978 & 0.4981 \\
		& GHE (q=1) 	 & 0.8607 & 0.8069 & 0.7562 & 0.7075 & 0.6636 & 0.6236 & 0.5882 & 0.5551 & 0.5261 & 0.5001 \\
		& GHE (q=2) 	 & 0.4999 & 0.4999 & 0.5000 & 0.5001 & 0.4999 & 0.5000 & 0.5001 & 0.4999 & 0.4999 & 0.5000 \\
\bottomrule
\end{tabular}
\caption{Expected values of $H$ for different $\alpha$s in the columns, with the simulations with the heaviest tails in the first column ($\alpha=1.1$) and the simulations with the normal distribution in the last column ($\alpha=2$). The results are also divided according to the different simulated time series lengths from $2^{9}$ to $2^{16}$ in the rows. Moreover, each row shows the results for the different methods of estimation ($R/S$, $DMA$, $MF-DFA$ with $q=1$, $DFA$ and $GHE$ for $q=1$ and $q=2$) and $\alpha$.}
\label{tab1}
\end{table}

\begin{table}[c]
\footnotesize
\begin{tabular}{@{ } l@{\hspace{4mm}} l@{\hspace{4mm}}c@{\hspace{2mm}}c@{\hspace{2mm}}c@{\hspace{2mm}}c@{ \hspace{2mm}}c@{\hspace{2mm}}c@{ \hspace{2mm}}c@{\hspace{2mm}}c@{\hspace{2mm}}c@{\hspace{2mm}}c@{\hspace{2mm}}c@{\hspace{2mm}}}
 \toprule \toprule
  & & \text{$\alpha $=1.1} & \text{$\alpha $=1.2} & \text{$\alpha $=1.3} & \text{$\alpha $=1.4} & \text{$\alpha $=1.5} & \text{$\alpha $=1.6} & \text{$\alpha
$=1.7} & \text{$\alpha $=1.8} & \text{$\alpha $=1.9} & \text{$\alpha $=2} \\
 \midrule \midrule
 $2^{9}$ 	& R/S    		 & 0.4172 & 0.4129 & 0.4058 & 0.4143 & 0.4058 & 0.4055 & 0.4107 & 0.4120 & 0.3934 & 0.4088 \\
  		& DMA 		 & 0.0877 & 0.1746 & 0.2124 & 0.2472 & 0.2390 & 0.2649 & 0.2514 & 0.2876 & 0.3009 & 0.3054 \\
  		& MF-DFA		 & 0.4533 & 0.4123 & 0.3931 & 0.3753 & 0.3818 & 0.3454 & 0.3492 & 0.3335 & 0.3091 & 0.3059 \\
		& DFA 		 & 0.2234 & 0.2414 & 0.2279 & 0.2564 & 0.2570 & 0.2477 & 0.2571 & 0.2713 & 0.2747 & 0.2910 \\
		& GHE (q=1) 	 & 0.6072 & 0.5834 & 0.5451 & 0.5106 & 0.4790 & 0.4580 & 0.4426 & 0.4233 & 0.4079 & 0.3850 \\
		& GHE (q=2) 	 & 0.3657 & 0.3899 & 0.3648 & 0.3847 & 0.3797 & 0.3814 & 0.3881 & 0.3858 & 0.3902 & 0.3920 \\		
 \midrule
 $2^{10}$ 	& R/S    		 &  0.4532 & 0.4450 & 0.4484 & 0.4400 & 0.4425 & 0.4411 & 0.4387 & 0.4462 & 0.4466 & 0.4350 \\
  		& DMA 		 & 0.2921 & 0.3181 & 0.3181 & 0.3063 & 0.3176 & 0.3720 & 0.3722 & 0.3861 & 0.3823 & 0.3698 \\
  		& MF-DFA		 & 0.5124 & 0.4829 & 0.4624 & 0.4456 & 0.4354 & 0.4125 & 0.3964 & 0.3933 & 0.3785 & 0.3565 \\
		& DFA 		 &  0.2711 & 0.2831 & 0.2682 & 0.3023 & 0.3007 & 0.3347 & 0.3361 & 0.3346 & 0.3432 & 0.3473 \\
		& GHE (q=1) 	 &  0.6680 & 0.6201 & 0.5868 & 0.5532 & 0.5276 & 0.5048 & 0.4787 & 0.4590 & 0.4436 & 0.4267 \\
		& GHE (q=2) 	 & 0.4126 & 0.4171 & 0.4165 & 0.4152 & 0.4141 & 0.4204 & 0.4195 & 0.4205 & 0.4272 & 0.4260 \\
 \midrule
 $2^{11}$ & R/S    		 & 0.4670 & 0.4622 & 0.4656 & 0.4670 & 0.4632 & 0.4668 & 0.4667 & 0.4640 & 0.4606 & 0.4671 \\
  		& DMA 		 & 0.3583 & 0.4014 & 0.4024 & 0.4207 & 0.4292 & 0.3988 & 0.4339 & 0.4166 & 0.4245 & 0.4243 \\
  		& MF-DFA		 & 0.5573 & 0.5310 & 0.5152 & 0.4866 & 0.4714 & 0.4547 & 0.4376 & 0.4215 & 0.4124 & 0.3935 \\
		& DFA 		 & 0.3532 & 0.3400 & 0.3658 & 0.3681 & 0.3721 & 0.3741 & 0.3813 & 0.3865 & 0.3860 & 0.3912 \\
		& GHE (q=1) 	 &  0.7086 & 0.6649 & 0.6321 & 0.5906 & 0.5619 & 0.5428 & 0.5120 & 0.4888 & 0.4648 & 0.4466 \\
		& GHE (q=2) 	 & 0.4459 & 0.4435 & 0.4458 & 0.4431 & 0.4425 & 0.4560 & 0.4473 & 0.4505 & 0.4486 & 0.4472 \\
 \midrule
 $2^{12}$ 	& R/S    		 & 0.4773 & 0.4725 & 0.4718 & 0.4760 & 0.4702 & 0.4747 & 0.4780 & 0.4763 & 0.4772 & 0.4733 \\
  		& DMA 		 &  0.4482 & 0.4445 & 0.4591 & 0.4553 & 0.4548 & 0.4532 & 0.4573 & 0.4567 & 0.4527 & 0.4555 \\
  		& MF-DFA		 & 0.5868 & 0.5687 & 0.5413 & 0.5075 & 0.4901 & 0.4766 & 0.4612 & 0.4519 & 0.4372 & 0.4239 \\
		& DFA 		 & 0.3671 & 0.3721 & 0.3800 & 0.3856 & 0.3845 & 0.3943 & 0.4070 & 0.4106 & 0.4064 & 0.4139 \\
		& GHE (q=1) 	 & 0.7321 & 0.6917 & 0.6535 & 0.6226 & 0.5884 & 0.5564 & 0.5308 & 0.5040 & 0.4827 & 0.4639 \\
		& GHE (q=2) 	 & 0.4618 & 0.4696 & 0.4537 & 0.4674 & 0.4669 & 0.4593 & 0.4674 & 0.4675 & 0.4647 & 0.4633 \\
 \midrule
 $2^{13}$ 	& R/S    		 &  0.4798 & 0.4799 & 0.4776 & 0.4819 & 0.4792 & 0.4823 & 0.4831 & 0.4803 & 0.4826 & 0.4870 \\
  		& DMA 		 & 0.4822 & 0.4834 & 0.4820 & 0.4787 & 0.4776 & 0.4776 & 0.4772 & 0.4821 & 0.4797 & 0.4796 \\
  		& MF-DFA		 & 0.6086 & 0.5773 & 0.5590 & 0.5348 & 0.5136 & 0.4890 & 0.4772 & 0.4666 & 0.4482 & 0.4382 \\
		& DFA 		 & 0.4029 & 0.3945 & 0.3975 & 0.4107 & 0.4153 & 0.4060 & 0.4186 & 0.4255 & 0.4303 & 0.4320 \\
		& GHE (q=1) 	 & 0.7587 & 0.7188 & 0.6792 & 0.6417 & 0.6050 & 0.5737 & 0.5446 & 0.5173 & 0.4943 & 0.4733 \\
		& GHE (q=2) 	 & 0.4801 & 0.4811 & 0.4750 & 0.4786 & 0.4791 & 0.4776 & 0.4751 & 0.4753 & 0.4748 & 0.4749 \\
 \midrule
 $2^{14}$ 	& R/S    		 &  0.4847 & 0.4834 & 0.4875 & 0.4872 & 0.4896 & 0.4850 & 0.4863 & 0.4880 & 0.4885 & 0.4898 \\
  		& DMA 		 & 0.5013 & 0.4952 & 0.4983 & 0.4969 & 0.4967 & 0.4946 & 0.4950 & 0.4970 & 0.4939 & 0.4934 \\
  		& MF-DFA		 & 0.6378 & 0.6083 & 0.5773 & 0.5503 & 0.5328 & 0.5138 & 0.4939 & 0.4773 & 0.4613 & 0.4495 \\
		& DFA 		 & 0.4135 & 0.4062 & 0.4072 & 0.4190 & 0.4362 & 0.4345 & 0.4396 & 0.4420 & 0.4395 & 0.4437 \\
		& GHE (q=1) 	 & 0.7862 & 0.7408 & 0.6937 & 0.6564 & 0.6190 & 0.5874 & 0.5546 & 0.5293 & 0.5039 & 0.4820 \\
		& GHE (q=2) 	 & 0.4881 & 0.4862 & 0.4834 & 0.4846 & 0.4840 & 0.4847 & 0.4855 & 0.4824 & 0.4828 & 0.4827 \\
 \midrule
 $2^{15}$ 	& R/S    		 & 0.4862 & 0.4867 & 0.4880 & 0.4874 & 0.4871 & 0.4863 & 0.4907 & 0.4939 & 0.4923 & 0.4924 \\
  		& DMA 		 & 0.5148 & 0.5107 & 0.5068 & 0.5013 & 0.5063 & 0.5048 & 0.5041 & 0.5052 & 0.5040 & 0.5039 \\
  		& MF-DFA		 & 0.6565 & 0.6233 & 0.5921 & 0.5683 & 0.5418 & 0.5202 & 0.5065 & 0.4893 & 0.4707 & 0.4582 \\
		& DFA 		 & 0.4239 & 0.4350 & 0.4344 & 0.4357 & 0.4378 & 0.4425 & 0.4462 & 0.4498 & 0.4489 & 0.4545 \\
		& GHE (q=1) 	 & 0.7983 & 0.7543 & 0.7072 & 0.6658 & 0.6288 & 0.5956 & 0.5639 & 0.5362 & 0.5105 & 0.4878 \\
		& GHE (q=2) 	 & 0.4933 & 0.4929 & 0.4915 & 0.4896 & 0.4886 & 0.4898 & 0.4891 & 0.4883 & 0.4883 & 0.4878 \\
 \midrule
 $2^{16}$ 	& R/S    		 & 0.4904 & 0.4890 & 0.4904 & 0.4904 & 0.4909 & 0.4888 & 0.4932 & 0.4920 & 0.4939 & 0.4956 \\
  		& DMA 		 & 0.5192 & 0.5151 & 0.5147 & 0.5152 & 0.5126 & 0.5137 & 0.5088 & 0.5127 & 0.5119 & 0.5092 \\
  		& MF-DFA		 & 0.6752 & 0.6411 & 0.6163 & 0.5809 & 0.5565 & 0.5332 & 0.5155 & 0.4965 & 0.4772 & 0.4659 \\
		& DFA 		 &  0.4301 & 0.4412 & 0.4365 & 0.4476 & 0.4409 & 0.4530 & 0.4517 & 0.4601 & 0.4618 & 0.4630 \\
		& GHE (q=1) 	 & 0.8134 & 0.7645 & 0.7186 & 0.6763 & 0.6376 & 0.6030 & 0.5702 & 0.5407 & 0.5143 & 0.4910 \\
		& GHE (q=2) 	 & .4963 & 0.4955 & 0.4939 & 0.4937 & 0.4919 & 0.4931 & 0.4927 & 0.4919 & 0.4909 & 0.4915 \\
\bottomrule
\end{tabular}
\caption{2.5\% Quantile values for different $\alpha$s in the columns, with the simulations with the heaviest tails in the first column ($\alpha=1.1$) and the simulations with the normal distribution in the last column ($\alpha=2$). The results are also divided according to the different simulated time series lengths from $2^{9}$ to $2^{16}$ in the rows. Moreover, each row shows the results for the different methods of estimation ($R/S$, $DMA$, $MF-DFA$ with $q=1$, $DFA$ and $GHE$ for $q=1$ and $q=2$) and $\alpha$.}
\label{tab2}
\end{table}

\begin{table}[c]
\footnotesize
\begin{tabular}{@{ } l@{\hspace{4mm}} l@{\hspace{4mm}}c@{\hspace{2mm}}c@{\hspace{2mm}}c@{\hspace{2mm}}c@{ \hspace{2mm}}c@{\hspace{2mm}}c@{ \hspace{2mm}}c@{\hspace{2mm}}c@{\hspace{2mm}}c@{\hspace{2mm}}c@{\hspace{2mm}}c@{\hspace{2mm}}}
 \toprule \toprule
  & & \text{$\alpha $=1.1} & \text{$\alpha $=1.2} & \text{$\alpha $=1.3} & \text{$\alpha $=1.4} & \text{$\alpha $=1.5} & \text{$\alpha $=1.6} & \text{$\alpha
$=1.7} & \text{$\alpha $=1.8} & \text{$\alpha $=1.9} & \text{$\alpha $=2} \\
 \midrule \midrule
 $2^{9}$ 	& R/S    		 & 0.6502 & 0.6653 & 0.6631 & 0.6794 & 0.6903 & 0.6944 & 0.6998 & 0.7068 & 0.7180 & 0.7179 \\
  		& DMA 		 & 0.6792 & 0.6778 & 0.6843 & 0.6888 & 0.6744 & 0.6886 & 0.7019 & 0.6908 & 0.6872 & 0.6959 \\
  		& MF-DFA		 & 1.0230 & 0.9745 & 0.9534 & 0.9143 & 0.8995 & 0.8762 & 0.7957 & 0.7730 & 0.7198 & 0.7051 \\
		& DFA 		 & 0.8513 & 0.7790 & 0.7645 & 0.7696 & 0.7376 & 0.7231 & 0.7283 & 0.7428 & 0.7111 & 0.7280 \\
		& GHE (q=1) 	 & 0.9761 & 0.9307 & 0.8955 & 0.8595 & 0.7993 & 0.7782 & 0.7303 & 0.6737 & 0.6485 & 0.5993 \\
		& GHE (q=2) 	 & 0.5777 & 0.5780 & 0.5767 & 0.5831 & 0.5737 & 0.5825 & 0.5788 & 0.5905 & 0.5855 & 0.5926 \\
 \midrule
 $2^{10}$ 	& R/S    		 & 0.6052 & 0.6183 & 0.6252 & 0.6315 & 0.6331 & 0.6394 & 0.6440 & 0.6483 & 0.6481 & 0.6577 \\
  		& DMA 		 & 0.6280 & 0.6323 & 0.6404 & 0.6427 & 0.6378 & 0.6386 & 0.6485 & 0.6481 & 0.6436 & 0.6498 \\
  		& MF-DFA		 & 0.9922 & 0.9692 & 0.9484 & 0.9317 & 0.8858 & 0.8292 & 0.7617 & 0.7376 & 0.6855 & 0.6340 \\
		& DFA 		 & 0.7095 & 0.6971 & 0.6838 & 0.7048 & 0.6533 & 0.6474 & 0.6425 & 0.6424 & 0.6404 & 0.6315 \\
		& GHE (q=1) 	 & 0.9605 & 0.9424 & 0.9045 & 0.8359 & 0.8060 & 0.7435 & 0.7055 & 0.6453 & 0.6059 & 0.5696 \\
		& GHE (q=2) 	 & 0.5537 & 0.5558 & 0.5594 & 0.5586 & 0.5556 & 0.5594 & 0.5577 & 0.5601 & 0.5620 & 0.5668 \\
 \midrule
 $2^{11}$ & R/S    		 & 0.5756 & 0.5802 & 0.5915 & 0.5983 & 0.5979 & 0.6073 & 0.6049 & 0.6160 & 0.6277 & 0.6226 \\
  		& DMA 		 & 0.5940 & 0.6051 & 0.6175 & 0.6071 & 0.6052 & 0.6070 & 0.6075 & 0.6188 & 0.6160 & 0.6188 \\
  		& MF-DFA		 & 0.9877 & 0.9532 & 0.9290 & 0.8871 & 0.8407 & 0.8026 & 0.7508 & 0.7012 & 0.6397 & 0.5980 \\
		& DFA 		 & 0.6769 & 0.6613 & 0.6435 & 0.6348 & 0.6292 & 0.6102 & 0.6203 & 0.6130 & 0.6001 & 0.6024 \\
		& GHE (q=1) 	 & 0.9572 & 0.9169 & 0.8881 & 0.8210 & 0.7780 & 0.7198 & 0.6940 & 0.6451 & 0.5871 & 0.5535 \\
		& GHE (q=2) 	 & 0.5377 & 0.5364 & 0.5407 & 0.5369 & 0.5449 & 0.5424 & 0.5430 & 0.5461 & 0.5474 & 0.5498 \\
 \midrule
 $2^{12}$ 	& R/S    		 & 0.5585 & 0.5660 & 0.5683 & 0.5770 & 0.5775 & 0.5796 & 0.5847 & 0.5896 & 0.5907 & 0.5979 \\
  		& DMA 		 & 0.5764 & 0.5777 & 0.5745 & 0.5828 & 0.5837 & 0.5873 & 0.5842 & 0.5876 & 0.5889 & 0.5875 \\
  		& MF-DFA		 & 0.9856 & 0.9773 & 0.9088 & 0.8721 & 0.8221 & 0.7823 & 0.7564 & 0.7053 & 0.6242 & 0.5723 \\
		& DFA 		 & 0.6346 & 0.6236 & 0.6142 & 0.6209 & 0.5926 & 0.5865 & 0.5790 & 0.5869 & 0.5865 & 0.5740 \\
		& GHE (q=1) 	 & 0.9532 & 0.9257 & 0.8878 & 0.8076 & 0.7549 & 0.7270 & 0.6627 & 0.6103 & 0.5701 & 0.5354 \\
		& GHE (q=2) 	 & 0.5245 & 0.5255 & 0.5253 & 0.5285 & 0.5268 & 0.5293 & 0.5314 & 0.5327 & 0.5335 & 0.5342 \\
 \midrule
 $2^{13}$ 	& R/S    		 & 0.5501 & 0.5533 & 0.5580 & 0.5604 & 0.5621 & 0.5661 & 0.5696 & 0.5769 & 0.5755 & 0.5815 \\
  		& DMA 		 & 0.5577 & 0.5655 & 0.5664 & 0.5654 & 0.5692 & 0.5688 & 0.5730 & 0.5744 & 0.5683 & 0.5711 \\
  		& MF-DFA		 & 0.9733 & 0.9509 & 0.8844 & 0.8572 & 0.8075 & 0.8025 & 0.7426 & 0.6619 & 0.6164 & 0.5597 \\
		& DFA 		 & 0.6064 & 0.5982 & 0.5972 & 0.5817 & 0.5809 & 0.5738 & 0.5646 & 0.5680 & 0.5589 & 0.5596 \\
		& GHE (q=1) 	 & 0.9517 & 0.9080 & 0.8597 & 0.8150 & 0.7492 & 0.6908 & 0.6498 & 0.6050 & 0.5623 & 0.5253 \\
		& GHE (q=2) 	 & 0.5174 & 0.5186 & 0.5182 & 0.5177 & 0.5195 & 0.5208 & 0.5219 & 0.5221 & 0.5230 & 0.5244 \\
 \midrule
 $2^{14}$ 	& R/S    		 & 0.5392 & 0.5439 & 0.5458 & 0.5508 & 0.5534 & 0.5557 & 0.5581 & 0.5626 & 0.5659 & 0.5692 \\
  		& DMA 		 & 0.5474 & 0.5530 & 0.5541 & 0.5569 & 0.5559 & 0.5540 & 0.5566 & 0.5568 & 0.5596 & 0.5610 \\
  		& MF-DFA		 & 0.9621 & 0.9338 & 0.9017 & 0.8367 & 0.8193 & 0.7556 & 0.7038 & 0.6736 & 0.6053 & 0.5485 \\
		& DFA 		 &  0.6215 & 0.5874 & 0.5887 & 0.5653 & 0.5684 & 0.5528 & 0.5586 & 0.5535 & 0.5467 & 0.5513 \\
		& GHE (q=1) 	 & 0.9497 & 0.9187 & 0.8639 & 0.8047 & 0.7298 & 0.6871 & 0.6348 & 0.5917 & 0.5513 & 0.5171 \\
		& GHE (q=2) 	 & 0.5082 & 0.5134 & 0.5120 & 0.5118 & 0.5144 & 0.5153 & 0.5157 & 0.5175 & 0.5153 & 0.5174 \\
 \midrule
 $2^{15}$ 	& R/S    		 & 0.5322 & 0.5359 & 0.5385 & 0.5434 & 0.5445 & 0.5444 & 0.5481 & 0.5521 & 0.5545 & 0.5577 \\
  		& DMA 		 & 0.5418 & 0.5437 & 0.5444 & 0.5453 & 0.5472 & 0.5455 & 0.5461 & 0.5495 & 0.5499 & 0.5501 \\
  		& MF-DFA		 & 0.9702 & 0.9458 & 0.8909 & 0.8592 & 0.7828 & 0.7853 & 0.7184 & 0.6586 & 0.5879 & 0.5380 \\
		& DFA 		 & 0.5707 & 0.5703 & 0.5673 & 0.5554 & 0.5582 & 0.5520 & 0.5487 & 0.5419 & 0.5426 & 0.5361 \\
		& GHE (q=1) 	 & 0.9477 & 0.9053 & 0.8473 & 0.7747 & 0.7365 & 0.6690 & 0.6230 & 0.5852 & 0.5469 & 0.5126 \\
		& GHE (q=2) 	 & 0.5058 & 0.5081 & 0.5080 & 0.5086 & 0.5096 & 0.5096 & 0.5113 & 0.5115 & 0.5120 & 0.5118 \\
 \midrule
 $2^{16}$ 	& R/S    		 & 0.5291 & 0.5323 & 0.5337 & 0.5355 & 0.5385 & 0.5408 & 0.5422 & 0.5452 & 0.5477 & 0.5489 \\
  		& DMA 		 & 0.5358 & 0.5365 & 0.5378 & 0.5399 & 0.5410 & 0.5411 & 0.5419 & 0.5425 & 0.5431 & 0.5436 \\
  		& MF-DFA		 & 0.9499 & 0.9240 & 0.8766 & 0.8518 & 0.7873 & 0.7341 & 0.6979 & 0.6392 & 0.5992 & 0.5330 \\
		& DFA 		 & 0.5659 & 0.5483 & 0.5612 & 0.5549 & 0.5575 & 0.5458 & 0.5387 & 0.5393 & 0.5358 & 0.5361 \\
		& GHE (q=1) 	 & 0.9488 & 0.8981 & 0.8406 & 0.7709 & 0.7139 & 0.6677 & 0.6191 & 0.5766 & 0.5390 & 0.5085 \\
		& GHE (q=2) 	 & 0.5036 & 0.5043 & 0.5059 & 0.5064 & 0.5064 & 0.5069 & 0.5078 & 0.5074 & 0.5086 & 0.5082 \\
\bottomrule
\end{tabular}
\caption{97.5\% Quantile values for different $\alpha$s in the columns, with the simulations with the heaviest tails in the first column ($\alpha=1.1$) and the simulations with normal distribution in the last column ($\alpha=2$). The results are also divided according to the different simulated time series lengths from $2^{9}$ to $2^{16}$ in the rows. Moreover, each row shows the results for the different methods of estimation ($R/S$, $DMA$, $MF-DFA$ with $q=1$, $DFA$ and $GHE$ for $q=1$ and $q=2$) and $\alpha$.}
\label{tab3}
\end{table}

\end{document}